\documentclass[fleqn,12pt,twoside]{article}
\usepackage{espcrc1}
\usepackage{graphicx}

\title{ Charmonium near the deconfining transition on the lattice }

\author{
Takashi~Umeda\thanks{
    talk presented by T.~Umeda at PANIC 2002}\address{
    Center for Computational Physics, University of Tsukuba,
          Tsukuba 305-8577, Japan \vspace{-0.25cm} },
Hideo~Matsufuru\address{
    Yukawa Institute for Theoretical Physics, Kyoto University,
           Kyoto 606-8502, Japan \vspace{-0.25cm} },
Osamu~Miyamura\thanks{deceased}\address{
    Department of Physics, Hiroshima University,
           Higashi-Hiroshima 739-8626, Japan},
and Kouji~Nomura\addressmark}

\begin{document}

\maketitle

\begin{abstract}
We study the charmonium properties at finite temperature
using quenched lattice QCD simulations.
Although a simple potential model analysis indicates no bound state at 
$T>1.05T_c$, our analyses of the spatial correlation between quark and 
anti-quark and the spectral function
indicate that a bound-state-like structure may survive even above $T_c$.
\end{abstract}

\section{Introduction}

Investigation of the QCD phase transition at finite temperature and
density is one of most important subjects in particle and nuclear
physics.
To create the quark gluon plasma phase (QGP), in which quarks and
gluons are deconfined and the chiral symmetry is restored, 
the high energy heavy ion collision experiments have been performed
and the RHIC experiment at BNL is in progress.
The $J/\psi$ suppression has been considered as one of most important
signals of formation of QGP phase.
However, theoretical understanding of this phenomenon is still not
sufficient to draw a conclusive scenario.

We investigate the charmonium properties at finite temperature
using lattice QCD simulations which enable us to incorporate
the nonperturbative effects of QCD from the first principle.
In this paper, we report our results of three such approaches:
({\it a}) a nonrelativistic potential model analysis with the static
    quark potential measured in lattice QCD,
({\it b}) the spatial correlation between quark and antiquark
     \cite{umeda01}, and
({\it c}) direct reconstruction of the spectral function from lattice
   data of correlators using the maximum entropy method and the standard
   $\chi^2$ analysis \cite{umeda02}.
At present stage, all these works are done in the quenched approximation
(without dynamical quark effects).
To circumvent severe restriction of degrees of freedom in Euclidean
temporal direction, we employ the anisotropic lattice on which
temporal lattice spacing is finer than the spatial one
(for setup of anisotropic lattice, see Ref.~\cite{matsufuru01}).

The results of these analyses are summarized as follows.
Although the simple potential model analysis ({\it a}) indicate that
there is no bound state above $T_c$, the latter two analyses imply
that the structure of QCD plasma is much involved and persistent
collective modes may exist.

\section{Potential model analysis}

\begin{figure}[t]
\includegraphics[width=79mm]{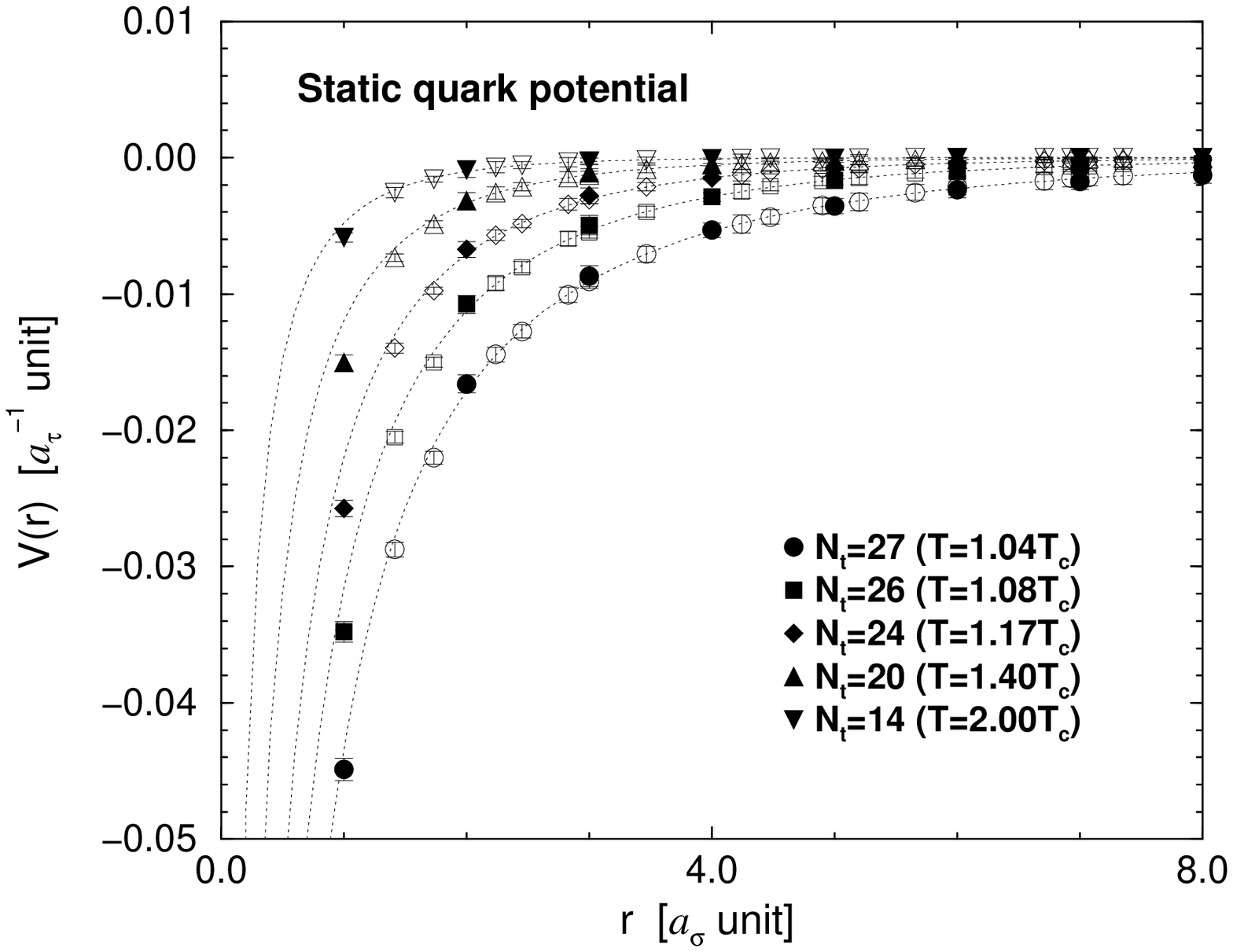}
\includegraphics[width=72mm]{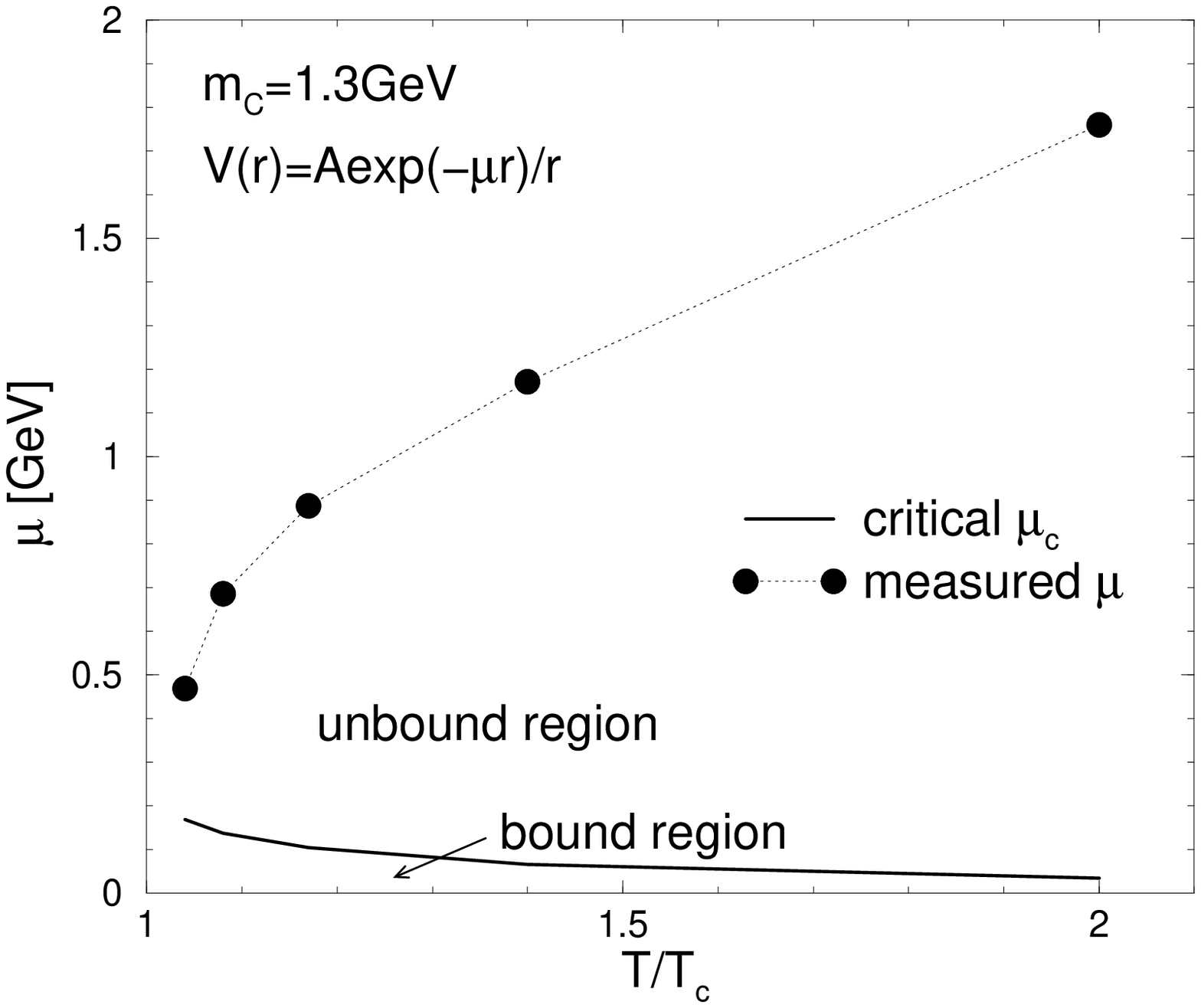}
\vspace{-1.0cm}
\caption{
The static quark potential above $T_c$ (left panel),
and the critical value $\mu_c$ and measured $\mu$ (right).}
\label{fig:potential}
\end{figure}

The static quark potential $V_{Q\bar{Q}}(\vec{r})$ is determined
in lattice QCD simulation.
It is extracted from the spatial correlation of the Polyakov loops,
$P(\vec{x})$:
\begin{equation}
 \langle P(0)P^\dagger(\vec{r})\rangle \simeq 
C \exp{(-V_{Q\bar{Q}}(\vec{r})N_t)},
  \hspace{0.4cm}
 P(\vec{x}) = \mbox{Tr}\, {\textstyle \prod_{t=0}^{N_t-1}} U_4(\vec{x},t).
\end{equation}
As displayed in the left panel of Fig.~\ref{fig:potential},
the results above $T_c$ are well fitted to the Yukawa-type form, 
$V_{Q\bar{Q}}(\vec{r})=-A\exp{(-\mu|\vec{r}|)}/r$, with the screening
length $1/\mu$.
The obtained value of $\mu$ at each temperature is shown in
the right panel of Fig.~\ref{fig:potential}.

To study whether $q$-$\bar{q}$ system has a bound state or not 
with the obtained potential, we solve the Schr\"odinger
equation for a spin averaged state, which is written for the
radial wave function for the S-wave, $\Phi(x)$, as
\begin{equation}
 \left( -\frac{\Delta}{2m_R}+V_{Q\bar{Q}}(x) \right)
 \phi(x)=E\phi(x),
\hspace{0.5cm} \phi(x)=x \Phi(x),
\end{equation}
where $m_R=m_c/2$ is the reduced mass and $m_c=1.3$ GeV is adopted. 
As the potential $V_{Q\bar{Q}}(x)$, we use the Yukawa-type
form and obtain the upper bound of the screening mass, $\mu_c$, 
below which the bound state is formed
(the solid line in Fig.~\ref{fig:potential}(right)).
Since the values of $\mu$  measured in the lattice simulation
are well larger than $\mu_c$,
we conclude that no bound state exists at $T> 1.05T_c$
from the the simple potential model analysis with the static
quark potential obtained in the quenched lattice QCD.
This result is consistent with Ref.~\cite{karsch88}.

\section{Spatial $q$-$\bar{q}$ correlation}

\begin{figure}[t]
\centerline{
\includegraphics[width=90mm,height=60mm]{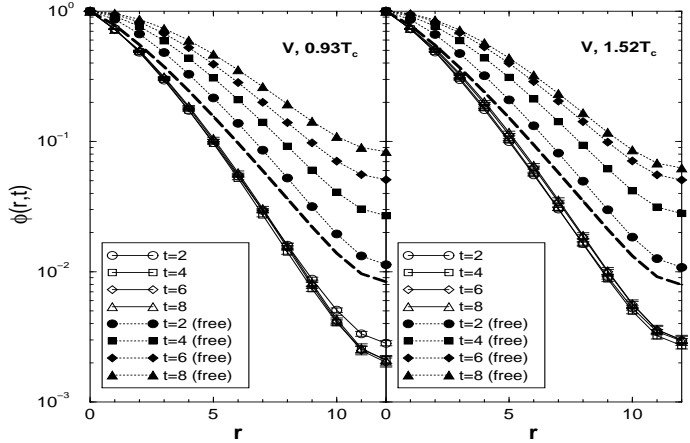}}
\vspace{-1.0cm}
\caption{
The spatial $q$-$\bar{q}$ correlation below and above $T_c$.
The dashed lines represent the smearing function.}
\label{fig:wavefunc}
\end{figure}

We examine a spatial $q$-$\bar{q}$ correlation (Bethe-Salpeter
amplitude) of the charmonium at finite temperature \cite{umeda01},
\begin{equation}
 w_\Gamma(r,t)={\textstyle \sum_{\vec{x}}}\,
             \langle\bar{q}(\vec{x}+\vec{r},t)\Gamma
             q(\vec{x},t)O^\dagger(0)\rangle,~~~
 O(\vec{x},t)= {\textstyle \sum_{\vec{y}}}\,
     \phi(\vec{y}) \bar{q} ( \vec{x} + \vec{y},t) \Gamma q(\vec{x},t),
\label{eq:BSf}
\end{equation}
where $\Gamma = \gamma_5$ for PS and $\gamma_i$ for V channels.
$\phi(\vec{y})$ is a smearing function for which we use the
result of $w_\Gamma(\vec{r},t)$ at $t\gg 1$ at $T\!=\!0$.
Figure~\ref{fig:wavefunc} shows the result of $q$-$\bar{q}$
correlation in the vector channel normalized at the spatial origin,
$\phi_\Gamma(r,t)=w_\Gamma(r,t)/w_\Gamma(0,t)$, together
with the cases of free quark and antiquark.
The $t$-dependence $\phi_\Gamma(r,t)$ give us a hint on the
behavior of quark and antiquark in plasma phase, since
if there is no bound state this spatial correlation becomes 
broader as $t$ increases, like the case of free quarks.
Contrary to naive expectation, even at $T\simeq 1.5T_c$ we observe
strong spatial correlation similar to that below $T_c$,
which is stable against $t$.
This implies that the quark and antiquark behave differently from
the free quarks, and still tend to stay together
up to this temperature.

\section{Spectral function}

The spectral function contains the most direct information
on the structure of the mesonic correlator,
$C(t)=\sum_{\vec{x}} \langle O(\vec{x},t) O^{\dag}(0,0) \rangle$.
Recently the maximum entropy method (MEM) has been successfully
applied to the extraction of the spectral function from lattice
data at $T=0$ without assuming specific form \cite{nakahara99}.
However, its applicability to problems at $T>0$ is not straightforward,
because available physical range of $C(t)$ as well as number of
data points are severely limited.
We propose to use a standard $\chi^2$ fit analysis in combination
with MEM, since the former gives more quantitative result for
the parameters such as mass and width than the latter if presumable
form of the spectral function is given, which can be estimated with MEM.
The numerical simulation is performed on lattices with spatial cutoff
$a_\sigma^{-1}\simeq 2$ GeV and the anisotropy $a_\sigma/a_\tau=4$.

To verify the applicability of these methods at $T>0$,
we require that they correctly work for the correlators at $T=0$
with restricted number of data points corresponding to the case
at $T>0$.
Unfortunately, MEM does not pass this test for the correlators
between local operators with present level of statistics
($O(500)$ configurations):
MEM fails to reproduce the correct feature
of the spectral function at low energy even qualitatively
for the region of $C(t)$ less than $O(0.5 \mbox{fm})$.
The reason may be that one need to extract the low energy structure
from the correlator in a short distance, where the correlator contains
the contribution from wide range of frequency of the spectral function.
This result warns against physical significance of MEM analysis
of such correlators at $T>0$.

To circumvent this problem, we apply the smearing technique
which enhances the low energy part of the spectral function.
For the correlators smeared with the wave function at $T=0$,
both the procedures works well in the low energy region.
Although the features of the collective mode such as the mass and
width are unchanged by the smearing,
the smearing can produce a mimic peak structure in the spectral
function \cite{wetzorke02}. 
In order to examine this possibility we also observe the correlators
smeared with a narrower function ({\it half-smeared} correlators).
The dependence of extracted spectral function on the smearing
function is compared with that of correlators composed of free
quarks.

\begin{figure}[t]
\centerline{
\includegraphics[width=90mm]{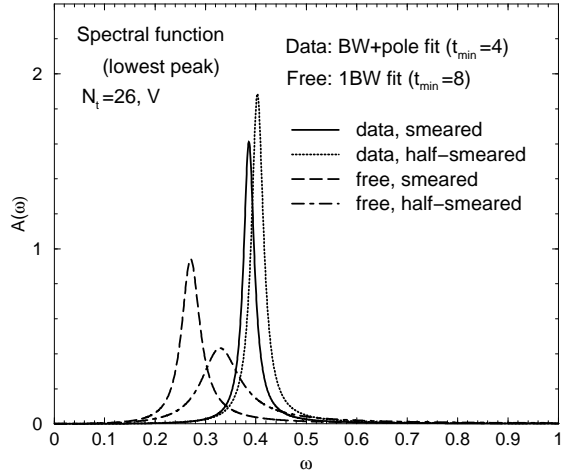}}
\vspace{-1.1cm}
\caption{
The spectral functions from the $\chi^2$ fit analysis
at $T\simeq 1.1T_c$.}
\label{fig:SPFs}
\end{figure}

At $T\simeq 0.9T_c$ the results indicate that the structure of the
ground state is almost the same as at $T=0$: almost the same mass
with the width consistent with zero.

At $T\simeq 1.1T_c$ the result of MEM still exhibits a peak structure 
around the same energy region as at $T<T_c$.
Therefore we perform the same kind of the $\chi^2$ fit analysis as
at $T<T_c$.
The fit to the 2-pole {\it ansatz} and to the relativistic
Breit-Wigner (BW) type {\it ans\"atze} give
inconsistent results, and the latter fits indicate that the spectral
function has a peak with almost the same mass as at $T<T_c$ and 
the width of order of 200 MeV, as shown in Fig.~\ref{fig:SPFs}.
The result depends on the smearing function only slightly,
in contrast to the free quark case, and hence we conclude that
this is a physically significant structure.

\end{document}